# Elements of nonlinear analysis of information streams

## A.M. Hraivoronska, D.V. Lande

Institute for Information Recording of the NAS of Ukraine
dwlande@gmail.com

**Abstract**

This review considers methods of nonlinear dynamics to apply for analysis of time series corresponding to information streams on the Internet. In the main, these methods are based on correlation, fractal, multifractal, wavelet, and Fourier analysis. The article is dedicated to a detailed description of these approaches and interconnections among them. The methods and corresponding algorithms presented can be used for detecting key points in the dynamic of information processes; identifying periodicity, anomaly, self-similarity, and correlations; forecasting various information processes. The methods discussed can form the basis for detecting information attacks, campaigns, operations, and wars.

## Introduction

To study information streams on the Internet, i.e., the streams of messages published on websites, social networks, blogs, etc., up-to-date techniques should be applied, because known methods of generalizing information arrays (classification, phase integration, cluster analysis, etc.) are not always suitable for an adequate quantitative reflection of processes occurring in the information space [Lande, 2007].

Quantitative analysis of the dynamics of information streams generated on the Internet has become one of the most informative methods for investigating topicality of various themes in the information space. Diverse qualitative factors influence the dynamics, with many of them cannot be described accurately. However, the general nature of the time-dependent number of thematic publications on the Internet allows constructing mathematical models, studying, and forecasting. The observed time-dependent volume of information streams convincingly shows that the mechanisms of their generation and dissemination are associated with complex nonlinear processes. This article is dedicated to this particular subject.

## A formal description of the information stream

For information streams to be described formally, we introduce some general assumptions. First let us define an information stream [Dodonov, 2009]. This definition corresponds to the classical definition used in the information theory.

Consider a segment of the time axis $(t_0, t)$ such that $t_0 > t$. Assume that the number of documents published during this period equals to $k$, which occurs according to some patterns. Suppose documents are published at times $\tau_1, \tau_2, \ldots, \tau_k$ ($t_0 \leq \tau_1 < \tau_2 < \ldots < \tau_k \leq t$).

The process $N_{t_0}(t)$ is called the information stream if the realization of this process consists of the numbers of points (documents) appeared from $t_0$ to $t$. Further, we are going to consider discrete time series corresponding to the function $N_{t_0}(t)$. Note that the values of the function $N_{t_0}(t)$ belong to $\mathbb{Z}$ and do not decrease; therefore, this function is a nondecreasing step function. The values of the time series are the step sizes. Further, we deal with discrete time series, with the values fixed at equidistant time intervals. Denote such time series by $x_1, x_2, \ldots, x_T$ or for short $\{x_t\}_{t=1}^T$, assuming that the observations are made at time interval $h$: $t_0, t_0 + h, t_0 + 2h, \ldots, t_0 + (T-1)h$.

The problem under consideration requires the time series to be described only in terms of the probability distribution, i.e., we consider statistical time series. In the article, we analyze a time series as a realization of some stochastic process.

As examples, we will use three time series obtained from the network service Google Trends. These time series reflect the level of interest to Donald Trump, Hilary Clinton and the information attacks of "Russian hackers" from August 2016 to April 2017. Time series obtained from Google Trends show the dynamics of the popularity for the search query. The maximum of the graph is 100, which corresponds to the date when the query was most popular, and all other points are determined as a percentage of the maximum. All the time series are shown in Figure 1. For referring conveniently, we denote these time series by T (D. **T**ramp), C (H. **C**linton), and H ("Russian **h**ackers")

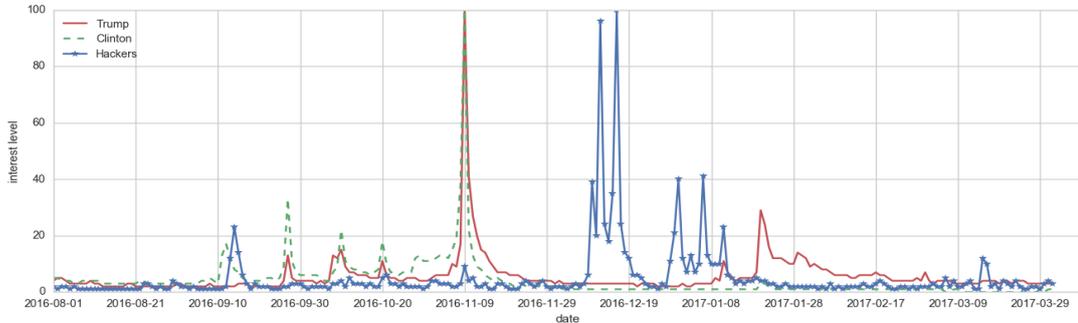

**Figure 1. Time series obtained from Google Trends. These time series reflect the level of interest to D. Trump, H. Clinton and "Russian hackers" from August 2016 to April 2017.**

### Correlation

Many methods of studying time series are based on some assumption about statistical equilibrium or stability. One of such useful assumptions is stationarity [Box, 2015].

A time series is called ***strictly stationary*** if its statistical properties are not affected by a change of time origin. Formally, if the joint probability distribution of random variables $x_t, x_{t+1}, \ldots, x_{t+n}$ is exactly the same as the joint probability distribution of $x_{t+k}, x_{t+k+1}, \ldots, x_{t+k+n}$ for all integer k, then the time series $\{x_t\}_{t=1}^T$ is said to be strictly stationary. A stationary time series has a constant mean and a constant variance

$$\mu = Ex_t, \ \sigma^2 = Var(x_t) = E(x_t - Ex_t)^2,$$

with the values $\mu$ and $\sigma^2$ can be estimated as the sample mean and the sample variance of the time series

$$\hat{\mu} = \bar{x} = \frac{1}{T} \sum_{t=1}^T x_t \, , \ \hat{\sigma}^2 = \frac{1}{T} \sum_{t=1}^T (x_t - \bar{x})^2. \tag{1}$$

The stationarity assumption is also of great importance for comparing time series. A measure of linear dependence between random variables is covariance. Defined for time series is the covariance function. By definition, ***cross-covariance at lag k*** *between random processes* $\{x_t\}_{t=1}^T$ *and* $\{y_t\}_{t=1}^T$ *is equal to*

$$\gamma_{xy}(k, t) = Cov(x_t, y_{t+k}) = E\big[(x_t - \mu_x)(y_{t+k} - \mu_y)\big].$$

It follows from the assumption of strict stationarity that the joint probability distribution of $x_t, \ y_{t+k}$ is the same for any $t$. It implies that the covariance between $x_t$ and $y_{t+k}$ do not depend on $t$, depending only on $k$. In other words, $\gamma_{xy}(k, t) = \gamma_{xy}(k)$ for all $t$. The collection of values $\{\gamma_{xy}(k)\}$ is called the cross-covariance function.

To get a cross-correlation coefficient, we normalize a cross-covariance coefficient

$$\rho_{xy}(k) = \frac{Cov(x_t, y_{t+k})}{\sigma_x \sigma_y} = \frac{\gamma_{xy}(k)}{\sigma_x \sigma_y}.$$

Note that the cross-correlation function is a measure of similarity between time series.

The cross-correlation and cross-covariance coefficients have following wide-used estimates [Montgomery, 2008]

$$\hat{\gamma}_{xy}(k) = \frac{1}{T}\sum_{t=1}^{T-k}(x_t - \bar{x})(y_{t+k} - \bar{y}), \qquad \hat{\rho}_{xy}(k) = \frac{\hat{\gamma}_{xy}(k)}{\hat{\gamma}_{xy}(0)}.$$

Figure 2 is provided to illustrate the definition of correlation. We consider two time series with zero means. To get the correlation coefficient between these time series we should multiply corresponding values and calculate their average. The result of the multiplication is shown as the line without markers in Figure 2. The area of the darkened region is equal to the covariance coefficient between the time series, taking into account the sign.

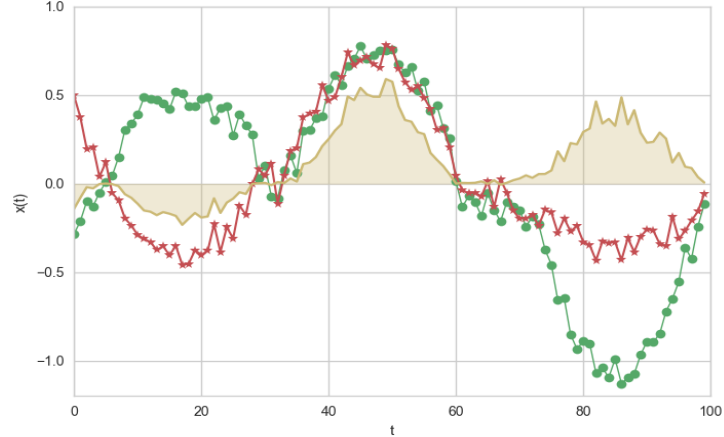

**Figure 2. The darkened region shows the contribution to the correlation between time series.**

Consider the estimation of the cross-correlation function for time series T and C (Figure 3). The function has maximum (approximately 0.8) at lag 0. It means that two time series which correspond to the interest level to Donald Trump and Hilary Clinton are strongly correlated.

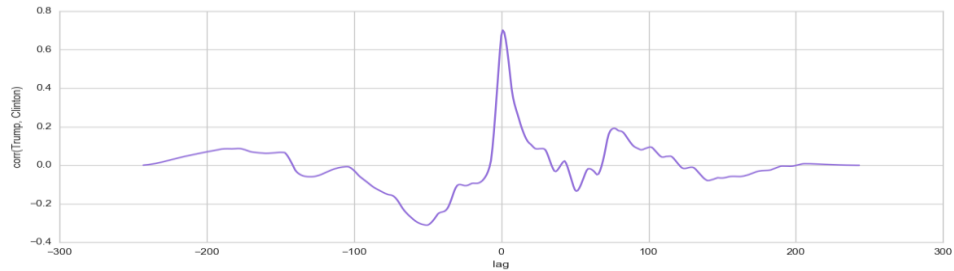

**Figure 3. The cross-correlation function for the T and C time series versus lag.**

## Autocorrelation

One can calculate the covariance not only using two different time series but also using just one time series. The covariance between two values of one time series is called the autocovariance at lag $k$

$$\gamma_k = Cov(x_t, x_{t+k}) = E[(x_t - \mu)(x_{t+k} - \mu)].$$

The collection of the values $\gamma_k,\ k = 0,1,2, \dots$ is called the autocovariance function and the collection of the normalized values $\rho_k,\ k = 0,1,2, \dots$ is called **_autocorrelation function_**

$$\rho_k = \frac{E[(x_t - \mu)(x_{t+k} - \mu)]}{\sqrt{E(x_t - \mu)^2 E(x_{t+k} - \mu)^2}} = \frac{Cov(x_t, x_{t+k})}{Var(x_t)} = \frac{\gamma_k}{\gamma_0}.$$

The autocorrelation function describes the correlation between values of a random process at different times [Chatfield, 2004].

In many cases the autocovariance and autocorrelation coefficients are estimated as follows

$$\hat{\gamma}_k = \frac{1}{T} \sum_{t=1}^{T-k} (x_t - \bar{x})(x_{t+k} - \bar{x}), \qquad \hat{\rho}_k = \frac{\hat{\gamma}_k}{\hat{\gamma}_0}.$$

Shown in Figure 4 is the autocorrelation function of time series T.

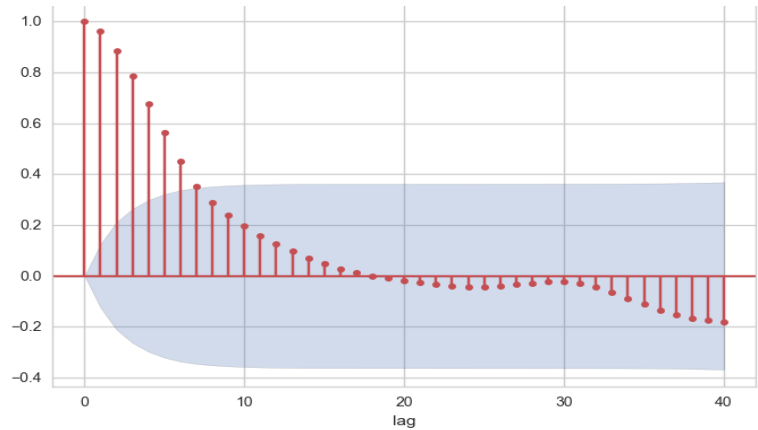

**Figure 4. The autocovariance function of the T time series versus lag.**

### Fourier analysis

The classical Fourier analysis is applied to study a function in the time and frequency domains. The idea behind the mapping to the frequency domain is that a function is decomposed into components that are harmonic oscillations at different frequencies. Each frequency corresponds to a coefficient that reflects the amplitude of the oscillation at a given frequency. If we represent the function graphically in the time domain, we get information about how the function changes over the time. If we represent the function in the frequency domain, we obtain information about the frequencies of the oscillations it contains. Mathematically speaking, to map a function from one domain to another the direct and inverse Fourier transform are used

$$\hat{x}(\nu) = \int_{-\infty}^{\infty} x(t) e^{-i2\pi\nu t} dt, \qquad x(t) = \int_{-\infty}^{\infty} \hat{x}(\nu) e^{i2\pi\nu t} d\nu.$$

Shown in Figure 5a is an example of the function defined as a sum of three sins with different periods. If we look at the graphical representation of the function in the time domain, it is quite difficult to understand that the function consists of three harmonic oscillations and determine their periods. Shown in Figure 5a is the Fourier transform for the function. From the graph in the frequency domain, it is clear that the function contains oscillations at three different frequencies.

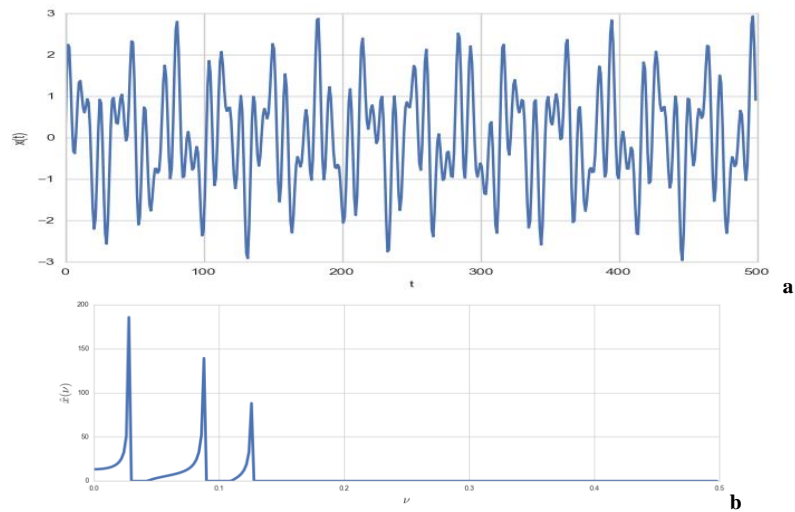

**Figure 5. (a) The sum of three sins with different periods.**
**(b) The estimated Fourier spectrum for the function shown in (a).**

The Fourier transform has various modern applications in machine learning systems. In many investigations, the Fourier spectra are used as train data. For example, proposed in [Rodrigues 2014] is a time series predictive model in which the Fourier spectrum with some other parameters is the input for a neural network.

The Fourier transform can be thought of as a correlation between the signal and different oscillating functions. Figure 6 illustrates this idea similarly to Figure 2.

Despite its advantages and numerous applications, the Fourier transform is a quite meaningless method for investigating functions that evolve over time. For such functions, we need some approach of estimating the spectrum not over the entire length of the time series, but over its various parts. An example of such an approach is Gabor transform

$$G(v, l, s) = \int\limits_{-\infty}^{\infty} x(t) e^{-\frac{(t-l)^2}{s^2}} e^{-i2\pi vt} dt.$$

The window function $e^{-\frac{(t-l)^2}{s^2}}$ gives higher weight to the part of a time series around the time $\tau$, with the parameter $s$ determining a width of the window.

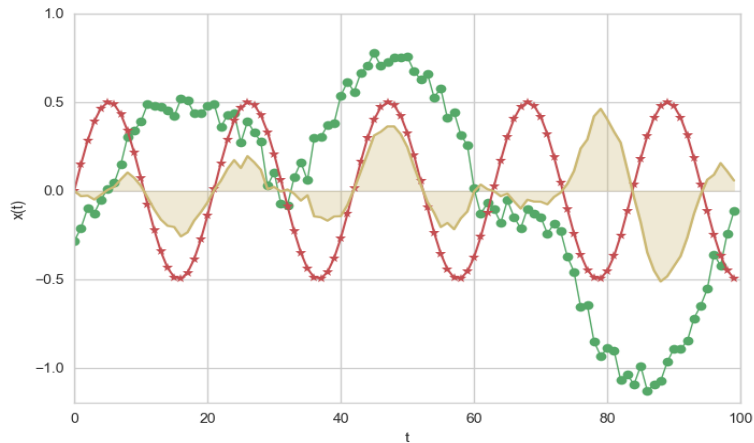

**Figure 6. An illustration to the definition of the Fourier transform as the correlation between the function in question and a harmonic function.**

When using the Gabor transform, there is a problem to choose the width of the window. The next type of transforms allows making the window function frequency-dependent in such a way that the window becomes wider for low frequencies and narrower for high ones. This type is called the wavelet transform. The main advantage of the wavelet transform is that a part of the time series is analyzed with the degree of detail that corresponds to its size.

### Wavelet analysis

The wavelet transform has a correlation nature. In this case, we consider a correlation of the function with a wavelet. For this procedure to be always possible to perform and the correlation coefficients to be informative, the wavelet must satisfy certain mathematical criteria. Literally, the word wavelet means a "small wave" and, as the name implies, the wavelet is well localized in time. Mathematically speaking, the wavelet is a function $\psi(t)$ which satisfies the following properties:

1.  The function $\psi(t)$ is square integrable ($\psi \in L^2(\mathbb{R})$) or, in other words, has finite energy

$$E = \int\limits_{-\infty}^{\infty} |\psi(t)|^2 dt < \infty.$$

2.  By $\hat{\psi}(\lambda)$ denote the Fourier transform of $\psi(t)$, then

$$\int\limits_{0}^{\infty} \frac{|\hat{\psi}(\lambda)|^2}{\lambda} d\lambda < \infty.$$

Shown in Figure 7 are examples of wavelets which are widely used in applications

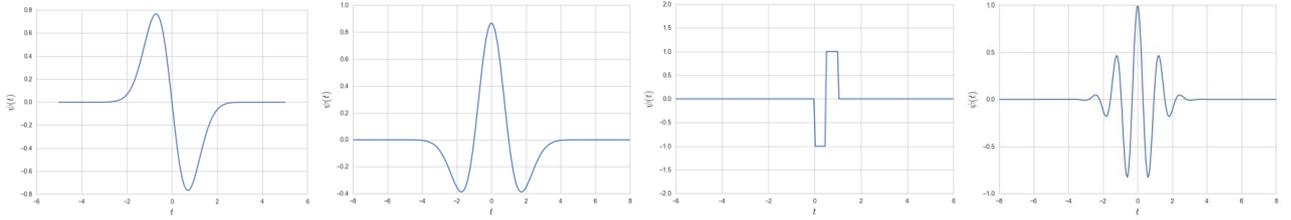

**Figure 7. Examples of wavelets to be applied in practice. The Gaussian wave, the Mexican hat wavelet, the Haar wavelet, and the Morlet wavelet (real part).**

## Continuous wavelet transform

The wavelet $\psi(t)$ described above is called the mother wavelet or analyzing wavelet. Based on the mother wavelet, a family of wavelets is constructed by stretching/squeezing and moving along the time axis. These transformations are necessary for exploring different parts of the original signal with various degrees of detail.

Let $s$ and $l$ be a scale parameter and a location parameter. Then the shifted and dialed version of the mother wavelet is as follows

$$\psi_{s,l}(t) = \frac{1}{\sqrt{|s|}} \psi\left(\frac{t-l}{s}\right).$$

The wavelet transform has been called a 'mathematical microscope,' where $l$ is the location of the time series being 'viewed' and $s$ is associated with the magnification at location b. [Addison, 2017]

***The continuous wavelet transform*** *of the function* $x(t) \in L^2(\mathbb{R})$ *is*

$$W(s,l) = \frac{1}{\sqrt{|s|}} \int\limits_{-\infty}^{\infty} x(t)\psi^*\left(\frac{t-l}{s}\right) dt = \int\limits_{-\infty}^{\infty} x(t)\psi_{s,l}^*(t) \, dt,$$

*where* $l, s \in \mathbb{R}$, $s \neq 0$; $\psi^*$ *is complex conjugate of* $\psi$, *the values* $\{W(s,l)\}_{l,s \in \mathbb{R}}$ *is called the coefficients of the wavelet transform.*

From the formula in the definition of the continuous wavelet transform, one can see that the essence of this transform is the calculation of correlation coefficients.

Let us visualize the wavelet coefficients for T, C, and H time series. Shown in Figure 8 are so called wavelet transform plots (plots of $W(s,l)$ versus $s$ and $l$). The results of the wavelet transform for T are shown in Figure 8 (a, b) where the Mexican hat wavelet (a) and the Gaussian wave wavelet (b) have been used. The results of the wavelet transform for C and H are shown in Figure 8 (c, d).

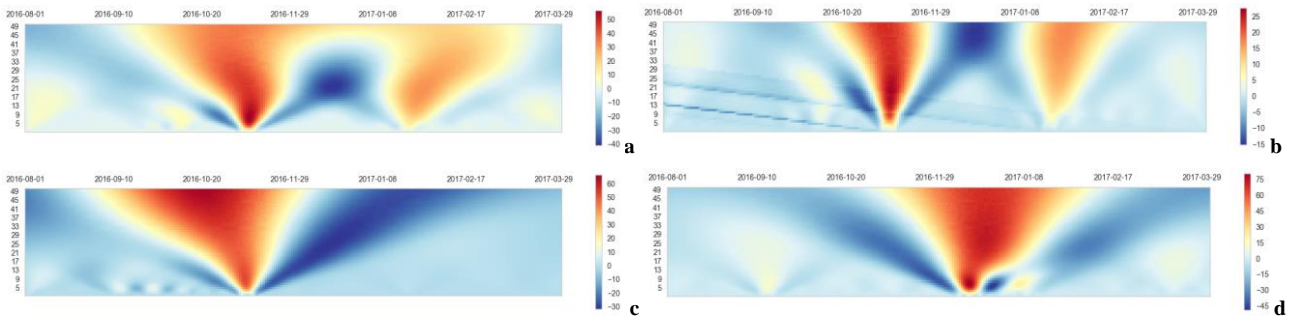

**Figure 8. (a, b) The wavelet transform coefficients for the time series T obtained using the Mexican hat wavelet and the Morlet wavelet; (c, d) The wavelet transform coefficients for the time series C and H obtained using the Mexican hat wavelet.**

Note once again that both the continuous Wavelet transform and the Fourier transform can be considered in terms of correlation. The Fourier transform is the correlation between the time series and the wave function $\varphi(t) = e^{-i2\pi\nu t}$. The wave covers the entire time axis and has the only one parameter $\nu$, therefore the Fourier transform depends only on the frequency. The wavelet transform is the correlation between the time series and the wavelet $\psi(t)$. Thus, the wavelet transform depends on the location of the wavelet on the time axis $l$ and its scale parameter $s$.

**Comparing time series using the wavelet transform**

Let us consider some approaches to comparing time series in wavelet space. These approaches can be applied to determine some kind of relationship between time series. Numerous metrics for comparing the coefficients of wavelet transforms as well as examples of their application to practical problems are described in detail in [Addison, 2017].

Consider two time series $x_t$ and $y_t$. By $W_x(s,l)$ and $W_y(s,l)$ denote their coefficients of the wavelet transform. We begin with the simplest way to compare two collections of values, namely, we substitute the absolute values of the coefficients

$$DiffMOD_{x,y}(s,l) = |W_x(s,l)| - |W_y(s,l)|.$$

Shown in Figure 9 are $DiffMOD_{x,y}(s,l)$ for pairs of time series: T and C, T and H. In this simple way, it is possible to highlight regions in which the wavelet coefficients are similar, and consequently, the time series have similar parts.

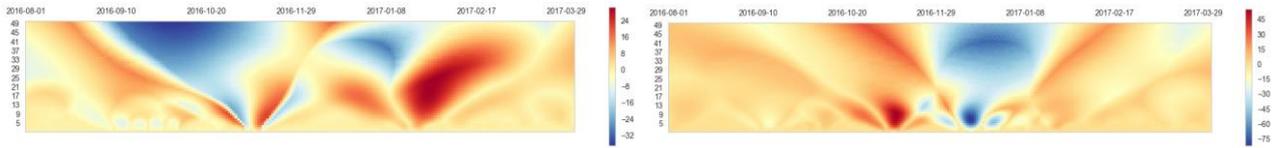

**Figure 9. The values of $DiffMOD_{x,y}(s,l)$ for the pair T and C (left) and the pair T and H (right).**

Additional information can be obtained by using a complex wavelet (for example, the Morlet wavelet). Then, in addition to the absolute value of the wavelet coefficients, one can use a phase. Recall that the complex coefficient can always be represented in the form

$$W(s,l) = |W(s,l)|e^{i\phi(s,l)}.$$

Therefore, we can compare the phases of the coefficients

$$\Delta\phi_{x,y}(s,l) = \phi_x(s,l) - \phi_y(s,l).$$

The cross-wavelet transform is used to determine regions of the coincidental energy between time series in the transform domain, as well as to get the relative phase

$$CrWT_{x,y}(s,l) = W_x^*(s,l)W_y(s,l).$$

The absolute values $|CrWT_{x,y}(s,l)|$ are often plotted to visualize this metric. In this case, the plot is similar to a scalogram, and if time series $x$ and $y$ are identical, we get exactly the scalogram.

Appling the cross-wavelet transform is of particular interest in the case when a complex wavelet is used. It follows

$$CrWT_{x,y}(s,l) = W_x^*(s,l)W_y(s,l) = |W_x(s,l)|e^{-i\phi_x(s,l)}|W_y(s,l)|e^{i\phi_y(s,l)} =$$
$$= |W_x(s,l)||W_y(s,l)|e^{i\left(\phi_y(s,l)-\phi_x(s,l)\right)}.$$

Thus, by calculating the cross-wavelet transform, it is possible to find the phase difference between the wavelet coefficients for two time series.

Shown in Figure 10 are $CrWT_{x,y}(s,l)$ for the T and C time series where the Mexican hat wavelet has been used. Highlighted for T and C (the left part of Figure 10) is the region corresponding to the peak of interest during the election. In this region both time series have high energy.

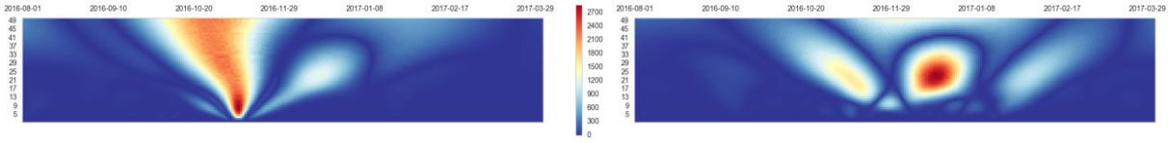

**Figure 10. The cross-wavelet transform for the pair T and C (left) and the pair T and H (right).**

The methods of cross-wavelet analysis are useful for studying properties of several time series with non-trivial dependence among themselves. For example, [Aguiar-Conraria, 2008] applies the cross-wavelet analysis tools to show that the relationship between monetary policy variables and macroeconomic variables has changed over time, and these changes are not homogeneous at different frequencies. Data obtained with the help of the cross-wavelet transform can also be used as input data for classification algorithms. In [Dey, 2010], the coefficients of the cross-wavelet transform are used as the input to an artificial neural network and a Fuzzy classifier.

### $\Delta L$-method

Scalograms obtained by the continuous wavelet transform are used to visualize special features of time series. In [Lande, 2009] another method of visualization is proposed. The method helps to identify trends, periodicity and local features of the time series. Moreover, the proposed approach is much easier to implement than the wavelet analysis.

The Δ-method is based on the DFA (Detrended Fluctuation Analysis) method, which will also be considered in the article. The essence of the approach is to determine and visualize the absolute deviation of the accumulated time series from the corresponding values of linear approximation.

Now we describe Δ-method in more detail. First, let us fix a length $s$ for a segment. We split up the time series into overlapping segments. For the point $x_t$ we choose the segment with length s and the center at the point $t$ (or at the point $t-1$ if s is even). For each segment fit the points in it with a linear function. Denote the value of local approximation at the point $j$ by $L_{t,j,s}$. Next, calculate the absolute deviation of $x_t$ from the approximation line, as follows $\Delta_{t,j,s} = |x_j - L_{t,j,s}|$.

According to the method we calculate values $\Delta_{t,j,s}$ for all $j = 1, \dots, T$ and $s = 1, \dots, [T/4\,]$. Firstly, we calculate standard deviation with given $s$

$$E(j,s) = \sqrt{\frac{1}{s}\sum_{t=1}^{T}|x_j - L_{t,j,s}|^2} = \sqrt{\frac{1}{s}\sum_{t=1}^{T}\Delta_{t,j,s}{}^2}.$$

Secondly, we take the average value of $E(j,s)$

$$F(s) = \frac{1}{T}\sum_{j=1}^{T}E(j,s).$$

Further, the values obtained are demonstrated on a diagram similar to the scaling diagram. An example of such diagram is shown in Figure 12.

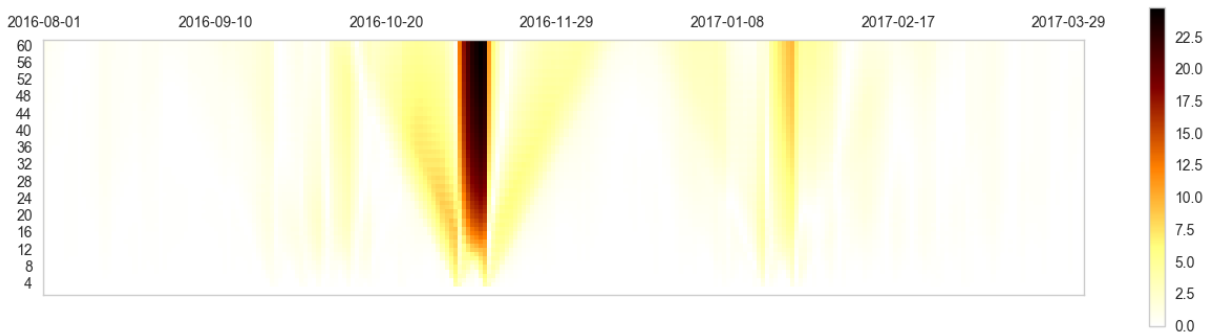

**Figure 12. The coefficients obtained with the $\Delta L$-method for T.**

The proposed method of visualizing absolute deviations $\Delta L$ allows to identify single and irregular peaks, sharp changes in the values of quantitative indicators at different times, as well as harmonic components in time series.

## Fractal analysis

Many objects in the world are statistically self-similar (a classic example is coastlines). Self-similarity means that parts of such objects have the same statistical characteristics at different scales. When studying the dynamic of information streams, the structure of document arrays on the Internet, and processes in the information space, self-similar structures often arise, and in particular, self-similar time series.

Let us define the self-similar process.

*The real-valued process $\{x(t), t \in \mathbb{R}\}$ is called **the self-similar process with the Hurst exponent $H > 0$** if the finite-dimensional distributions of $\{x(\alpha t), t \in \mathbb{R}\}$ are identical to the finite-dimensional distributions of $\{\alpha^H x(t), t \in \mathbb{R}\}$ for all $\alpha > 0$. This requirement may be written economically as*

$$\{x(\alpha t), t \in \mathbb{R}\} =^d \{\alpha^H x(t), t \in \mathbb{R}\}.$$

By definition, changing the time scale for a self-similar process is equivalent to changing the scale of the process values. This means that realizations of such processes are similar at different scales. However, it is natural that the process is not an exact copy of itself at different scales, only statistical properties are the same.

The Hurst exponent is known to be a measure of persistence - the tendency of the process to trends. The value H = 0.5 means that behavior of time series is uncorrelated, as in Brownian motion. Values in the range 0.5 <H <1 mean that the directed dynamics of the process in the past is likely to lead to proceeding the movement in the same direction. If H< 0.5 then the process is likely to change the direction [Braichevsky, 2010].

Let us describe some properties of self-similar processes that are of particular importance for applications. First, the autocovariance function of a self-similar process decays to zero hyperbolically, and the following estimate takes place

$$\rho_k \approx k^{(2H-2)} L(t) \; as \; k \to \infty,$$

where $L(t)$ is a slowly-varying-at-infinity function. Consequently, the series of covariance coefficients is divergent

$$\sum_{k=1}^{\infty} \rho_k = \infty.$$

This divergent series indicates a long-term dependence in the time series.

Second, let us split the time series $\{x_t\}$ into non-overlapping segments with $m$ elements in each one. Denote by $\left\{x_t^{(m)}\right\}$ the means for each segment. Then the variance of the mean decays slower than the inverse to the size of the sample $\sigma^2\left(x_t^{(m)}\right) \sim m^{2H-2}$.

## Estimating the Hurst exponent

The best-known method for estimating the Hurst exponent is rescaled range (R/S) method. Let us recall the standard deviation for a time series $\{x_t\}_{t=1}^T$

$$S = \sqrt{\frac{1}{T}\sum_{t=1}^{T}(x_t - \bar{x})^2}, \;\; где \; \bar{x} = \frac{1}{T}\sum_{t=1}^{T}x_t,$$

and the range

$$R = \max_{1 \le t \le T} x^{(t)} - \min_{1 \le t \le T} x^{(t)}, \text{где } x^{(t)} = \sum_{i=1}^{t}(x_i - \bar{x}).$$

Dividing R by S we get the rescaled range. For many time series observed the rescaled range is well-fitted by the following power law

$$\frac{R}{S} = \left(\frac{T}{2}\right)^{H}.$$

The estimate of the Hurst exponent can be obtained by calculating $R/S$ statistic as a function of $T$, plotting $\log R/S$ versus $\log T$, and fitting a straight line. The slope of the line gives the estimate of the Hurst exponent.

We will apply $R/S$ method to get the Hurst exponent for the T, C and H time series. Shown in Figure 12 are the results of assessing for T series. The obtained values of the Hurst exponent are 0.62 for T and 0.68 for C, which points out that these processes appear to follow a trend, although not very strictly.

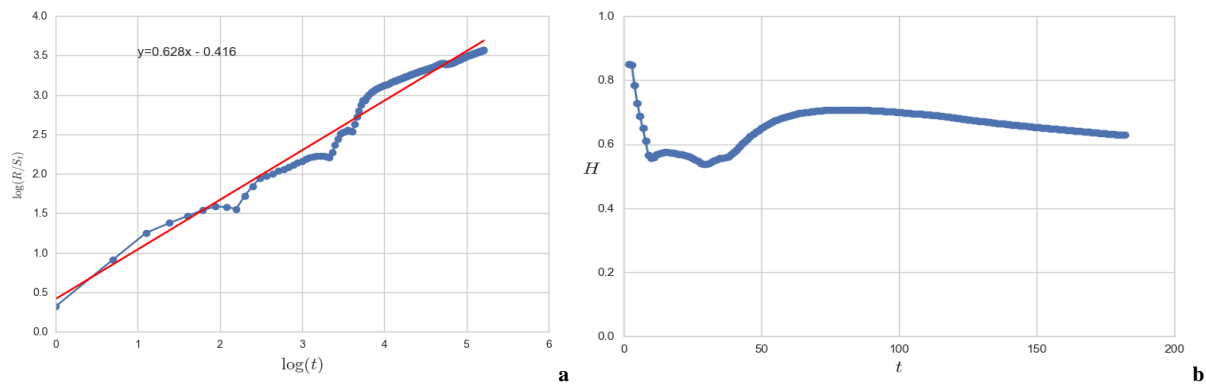

**Figure 12. (a) The dependence of the R / S statistics on time for the T series on a log-log plot. (b) The dependence of the Hurst exponent on time for the T series.**

As for the H time series, the dependence of $\log R/S$ on $\log t$ is poorly fitted by a straight line, because the line shown in Figure 13a has a strong fracture. If we plot the dependence of the Hurst exponent on time (Figure 13b), we can determine the time from which $H$ begins to decrease. Marking this time on the plot of H time series (Figure 13c), it becomes clear that the process before this time has much smaller variance than after.

The behavior of the H time series since the beginning of December 2016 (the beginning of the largest peak in the values of the time series) can be analyzed separately from the first part. The estimate of the Hurst exponent for the second part of the time series equals to 0.7. It should be taken into account that this time series becomes too short for $R/S$ analysis. Time series must have at least 200 elements and better more than 300 elements to achieve correct results. Nevertheless, the sharp change in the dependence of the Hurst exponent on time indicates that the process under investigation is made up of various processes. In such cases, it may be useful to analyze these processes separately.

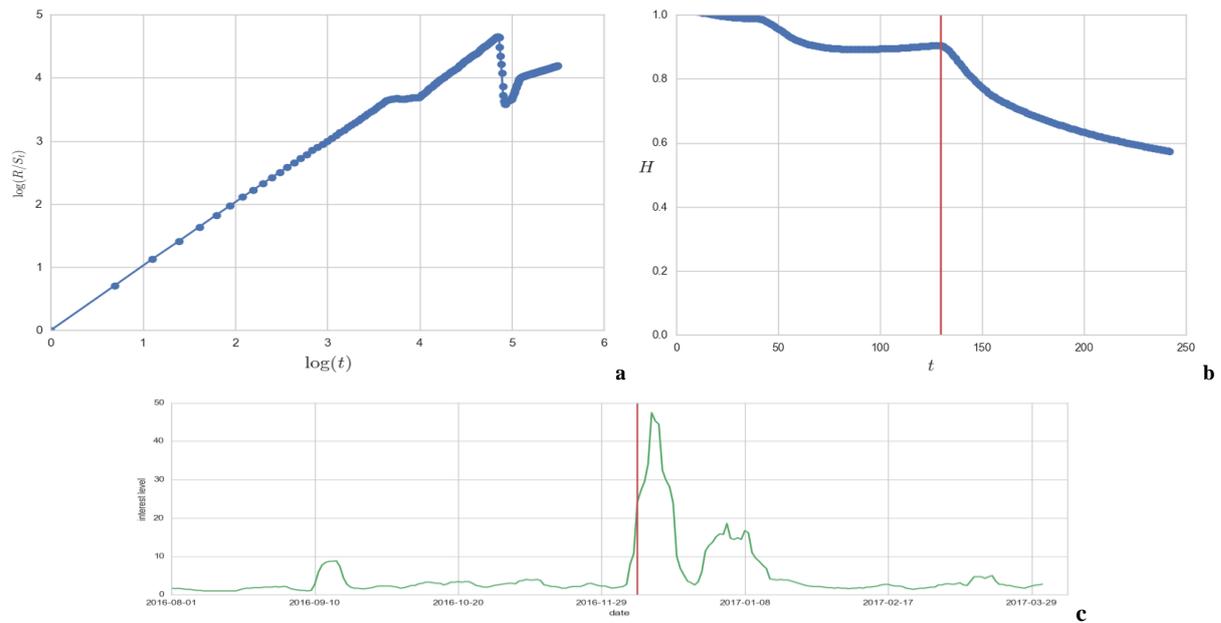

**Figure 13. (a) The dependence of the R / S statistics on time for the H series on a log-log plot. (b) The dependence of the Hurst exponent on time for the H series. (c) The H time series with the time of a rapid change marked.**

### Multifractals

The only one fractal dimension turns out to be not enough to describe many self-similar objects found. In many cases, real objects are not homogeneous. The multifractal theory provides a more general description of the essential characteristics of complex self-similar objects. According to this theory, an object is characterized by a continuous spectrum of dimensions, which makes possible to distinguish homogeneous objects from heterogeneous ones.

Multifractal set (signal) can be understood as a union of homogeneous fractal subsets (signals) with different fractal dimensions. The fractal dimensions are represented in a multifractal spectrum to be defined later. It is worth noticing that the multifractal spectrum can be used as a measure of similarity between objects. We can use such approach, for example, to form representative samples from document arrays, as an addition to traditional methods based on the identification of content similarity of documents. This approach has following practical applications: presenting to the user adequate number of search results that reflect the full range of the documentary array; selecting subsets of documents for further research [Lande, 2009a].

### The Hölder exponent and multifractal analysis

To define the multifractal spectrum, the Hölder exponent will be required. The Hölder exponent characterizes the smoothness of a function and contains information on the regularity of a function in a neighborhood of a point. The smaller the Hölder exponent is, the less regular the function is.

*Let $x$ be a locally bounded function on $\mathbb{R}$ and $t \in \mathbb{R}$.* ***The Hölder exponent*** *of $x$ at $t$*

$$h_x(t) = sup\{\alpha \geq 0 : |x(t + \Delta t) - x(t)| = O(\Delta t^\alpha), \Delta t \to 0\}.$$

In other words, the Hölder exponent characterizes local behavior of the function in the following way

$$|x(t + \Delta t) - x(t)| \sim C_t \Delta t^{h_x(t)}.$$

It is mentioned above that multifractal objects are not homogeneous; hence, their pointwise Hölder regularity changes from point to point. Therefore, it is useful to obtain information about sets in which the exponent takes a given value

$$E_h = \{t \in \mathbb{R} : h_x(t) = h\}.$$

To make some meaningful conclusions about $E_h$ one should determine and compare the sizes of such sets. It follows from practical cases that the fractal dimension is a reasonable formalization of size to be used in such analysis [Aldroubi, 2016]. This reasoning leads to the definition of the multifractal spectrum.

***The multifractal spectrum*** *of a locally bounded function* $x: \mathbb{R} \to \mathbb{R}$ *is the mapping*

$$d_x(h) = D_H(E_h)$$

Thus, the multifractal spectrum reveals the values of the Hölder exponent contained in the heterogeneous object (set, signal, measure) and proportionality among them. Each value of the Hölder exponent corresponds to the fractal dimension of the set in which the value of the Holder exponent is equal to a given one.

### An approach to estimating the multifractal spectrum

The theoretical approach to the determination of the multifractal spectrum is described above. For practical purposes, a calculation of the Hölder exponent at each point and determination of the fractal dimensions are numerically unstable and often meaningless. Therefore, a numerical approach is different. It based on the following lemma which follows from the definition of the Hölder exponent.

<u>Lemma.</u> Let x: $\mathbb{R} \to \mathbb{R}$ be a locally bounded function and $h_x(t) = H \in [0,1]$, then

$$h_x(t) = \lim_{j \to \infty} \inf \left[ \frac{log\left(R_x\left(B\left(t, 2^{-j}\right)\right)\right)}{log\, 2^{-j}} \right]$$

where denoted by $R_x(A) = max_{t \in A}\, x(t) - min_{t \in A}\, x(t)$ is the range of the function $x$ on the set $A$.

It follows from the lemma that the multifractal formalism for functions can be based on the structure function

$$Z(q,j) = \frac{1}{2^j} \sum_i R_x\left(B\left(\frac{i}{2^j}, \frac{1}{2^{j+1}}\right)\right)^q \qquad (2)$$

and the corresponding scale function

$$\tau(q) = \lim_{j \to \infty} \inf\left(\frac{log(Z(q,j))}{log\, 2^{-j}}\right), \qquad (3)$$

that leads to defining the multifractal spectrum as follows

$$d_x(h) = \inf_{q \in \mathbb{R}}(1 - \tau(q) + hq).$$

The expression for the multifractal spectrum through the scale function can be applied to the numerical analysis of time series. First, the structure function and the scale function are determined, and then through the Legendre transform, we come up with the multifractal spectrum [Mallat, 2009].

### The DFA method and its application to estimating the multifractal spectrum

[Peng, 1994] proposes the Detrended Fluctuation Analysis (DFA) method to determine long-term correlations in noisy and non-stationary time series. The key feature of the DFA method is that it is based on the theory of random walks. The time series is <u>not</u> analyzed in the original form. Instead, the series is transformed to accumulated sums

$$y_t = \sum_{k=1}^t x_k.$$

In this case, the point $y_t$ appears to be the location of the random walk after $t$ steps. Further, the DFA method involves an analysis of the mean square deviation from the trend in various regions of the series.

For the DFA method, many modifications have been proposed, as well as applications to numerous practical tasks. An overview of such methods is given, for example, in [Kantelhardt, 2009]. In particular, $\Delta L$-method introduced above is based on the DFA method. An important step is the development of an approach to the numerical estimation of the multifractal spectrum based on the DFA method. This method is called Multifractal Detrended Fluctuation Analysis (MF-DFA) and is proposed in [Kantelhardt, 2002]. The effectiveness of the MF-DFA method was analyzed for various model time series (Brownian motion, fractional Brownian motion, binomial cascades) [Oswiecimka, 2012]. The method is also actively used for the analysis of real time series, for example, economic ones [Suarez-Garcia, 2013]. A step-by-step description and explanation of the MF-DFA algorithm can be found in [Thompson, 2016].

### The use of wavelets for estimating the multifractal spectrum

Wavelets are a natural tool to investigate the fractal characteristics of an object [Jaffard, 2004] [Aldroubi, 2016]. The reason for this is the fact that the pointwise Hölder exponent can be estimated through the wavelet coefficients. The following proposition connects the wavelet transform with the Hölder exponent.

Proposition. *Let the function $x$ has the Hölder exponent $h_x(t)$ at the point $t$. Assume that the function $x$ at $t$ has not an oscillating singularity. If the first $n$ moments of the wavelet $\psi$ are equal $0$ and $n > h(t_0)$, then*

$$W_\psi[x](s,t) \sim s^{h_x(t)}, \quad s \to 0^+. \tag{4}$$

*Otherwise, if we choose the wavelet with $n < h(t_0)$, then*

$$W_\psi[x](s,t) \sim s^n, \quad s \to 0^+. \tag{5}$$

From the proposition it follows that the behavior of the function $x$ in a neighborhood of $t$ can be characterized as follows: the smoother the function is, the more quickly $W_\psi[x](s,t)$ decays as s tends to zero. For instance, if the function $x$ is continuously differentiable at the point $t$ (this means that $h_x(t) = +\infty$), then the wavelet transform approximately has the form (5). It means that the values $W_\psi[x](s,t)$ depend on the form of the wavelet. In practice, we often meet a fundamentally different case in which the function $x$ at the point $t$ has a Hölder exponent in the range from 0 to 1. In this case, the formula (4) is valid. This connection between the approximate behavior of wavelet coefficients and the Hölder exponent is of essential importance for the wavelet-based estimation of the multifractal spectrum.

### The use of maxima lines for estimating the spectrum. WTMM

The approach to estimating the multifractal spectrum using the maximum lines has been actively developed since [Mallat, 1992]. The expressions (4) and (5) turns out to be valid if we consider the curve of local modulus maxima of $W_\psi[x](s,t)$ instead of the sequence $W_\psi[x](s,t)$ for a constant $t_0$ and decreasing scale $s$. First, we introduce the necessary definitions.

*A modulus maximum is a point $(s_0, t_0)$ such that $W_\psi[x](s_0,t) \leq W_\psi[x](s_0,t_0)$, where denoted by $t$ is either left or right neighbor of $t_0$, with the strict inequality $W_\psi[x](s_0,t) < W_\psi[x](s_0,t_0)$ hold for at least one neighbor of $t_0$ (left or right). A maxima line is a connected curve in the scale space $(s,t)$ along which all points are modulus maxima. The set of maxima lines obtained from the wavelet transform of the function is called the skeleton.*

The skeleton for the series T is shown in Figure 15 (on the left, the skeleton is plotted over the wavelet coefficients). Such plots may be useful for visualization purposes.

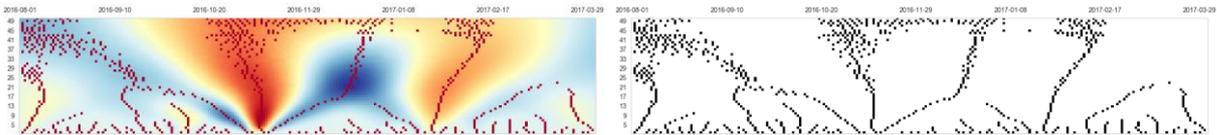

**Figure 15. The skeleton (the maxima lines) for the T series.**

The structure function depended on the maxima lines is as follows

$$Z(q,s) = \sum_{l \in \mathcal{L}(s)} \left( \sup_{\substack{(\tau,s') \in l \\ s' < s}} |W_\psi[x](t,s)| \right)^q, \qquad q \in \mathbb{R}.$$

Knowing the structure function, we can use formula (2) and (3) to determine the scaling function and the multifractal spectrum. Shown in Figure 16 are the scaling functions and the multifractal spectra for the T, C and H series. For comparing, Figure 16a shows the theoretic scaling function for Brownian motion $\tau(q) = q/2 - 1$.

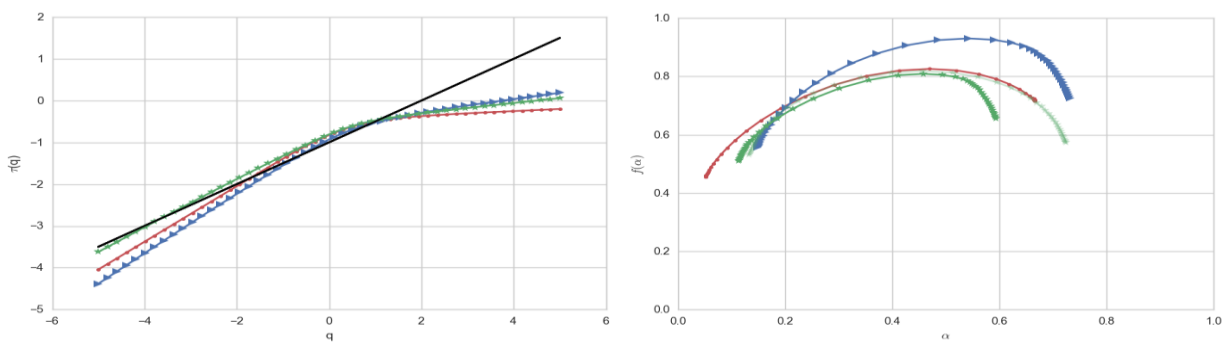

**Figure 16. The scaling function for the T (a circle), C (a triangular), and H (an asterisk) time series; and the corresponding multifractal spectra estimated with the WTMM method.**

The multifractal spectrum can be used to compare time series. The same shape of multifractal spectra indicate the similarity of the series, otherwise, the different shapes of the spectra mean that there are fundamental differences in the nature of the series. This property is used in various studies. For example, [Sun, 2001] analyzes statistically how varying certain parameters affects the multifractal spectrum for the Hang Seng index in the Hong Kong stock market. [Zhou, 2009] presents an overview of the scientific results in the field of economics based on a comparison of multifractal spectra. Using the Dow Jones Industrial Average as an example, the author identifies economic factors that affect the shape of the spectrum.

### Conclusion

Modern methods based on nonlinear analysis are to be applied for efficient analysis of modern information processes based on the monitoring of information streams from global computer networks, with many methods are of successful application in natural sciences. Up-to-date approaches to the analysis and modeling of public and information systems can be based on methods approved in natural sciences. Analyzing information streams is the foundation for modeling, design, and forecasting. The approaches discussed in this article allow describing information processes, information influence, general trends in the dynamics of information processes. At the same time, progress in mastering the modern information space is impossible without forming general ideas about the structure and properties of the dynamics of network information processes. This process requires identifying and taking into account stable patterns of the dynamics. Note that approaches based on the accurate methods, mathematical theory, and computer simulating can provide one-up qualitative results. This is conditioned by the multiparametericity of real models. Nevertheless, even such results can explain reality in many cases better than traditional qualitative methods.

The methods, algorithms, analytical tools discussed in this article are not only a demonstration basis for explaining real events and processes but also necessary components for planning and forecasting.